# Through-the-Wall Imaging Exploiting 2.4GHz Commodity Wi-Fi

Wei Zhong, Kai He, and Lianlin Li, *Senior Member, IEEE,*

*Abstract*—In this letter, we experimentally investigate a low-cost through-the-wall imaging exploiting Wi-Fi signals in an indoor environment from the perspective of holographic imaging. In our experiments, a pair of antennas in a synthetic aperture mode is used to acquire signals produced by commodity Wi-Fi devices and reflected from the scene in a synthetic aperture mode. The classical filtered back propagation (FBP) algorithm is then employed to form the image based on these signals. We use an IEEE 802.11n wireless router working at 2.4GHz with bandwidth of 20MHz. Selected experimental results are provided to demonstrate the performance of the proposed Wi-Fi based imaging scheme.

## I. Introduction

Through-the-wall imaging using radio frequency (RF) signals has become popular in diverse areas such as subsurface exploration, surveillance, life rescue operations, to name a few [1-8]. In such applications, the use of wideband or ultra-wideband (UWB) radar is highly desirable [2] to achieve high range resolution. However, conventional UWB radar techniques suffer from relatively expensive hardware deployment costs and most of through-the-wall systems operate in an active manner, which means that both transmitting and receiving functions must be taken into consideration.

With the increasing deployment of wireless local area networks (IEEE 802.11), electromagnetic signals based on this standard have become ubiquitous in many places, especially in urban areas and indoor environments. Numerous efforts have been made to explore Wi-Fi signals to achieve different low-cost indoor surveillance tasks. These include detection [6, 7] or location [8] of human subjects behind a wall, monitoring of human respiration [9], tracking the movement of people [10, 11], and body gesture recognition [12, 13]. These efforts can be mathematically classified as a *detection* problem. Recently, some efforts have been made to investigate the feasibility of Wi-Fi signals for *imaging* purposes, with encouraging results. This is despite of the fact that the bandwidth of Wi-Fi signals is relatively narrow (20MHz or 40MHz) compared with that of UWB radar, causing a relatively lower co-range resolution and accuracy [14-18]. For instance, Huang et al. [17] demonstrated that localization accuracy on the order of tens of centimeters can be achieved based on 2.4GHz Wi-Fi signals where an antenna array is used to acquire Wi-Fi signals scattered from the probed scene. In this case, however, the Wi-Fi signal is generated using a controlled universal software radio peripheral (USRP) [17] rather than commodity (off-the-shelf) Wi-Fi routers. Holl and Reinhard treated the Wi-Fi-based imaging from the perspective of holography, and verified experimentally the capability of 2.4GHz and 5GHz Wi-Fi signals for three-dimensional imaging [18].

Motivated by these advances, this work investigates the use of Wi-Fi signals based on commodity routers to perform through-the-wall imaging from the perspective of holographic imaging. In particular, our experiments perform data acquisition in a synthetic aperture mode, where two antennas receive the Wi-Fi signal, one as the scanning antenna and the other as the reference (fixed) antenna. We use a 2.4GHz IEEE 802.11n protocol wireless router with 20MHz bandwidth. The classical back-propagation algorithm is employed to process the calibrated Wi-Fi signal and produce three-dimensional images.

The remaining of this letter is organized as follows. Section II provides the Wi-Fi signal based imaging configuration and the data processing steps to yield the three-dimensional image. Section III reports on selected experimental results, including an imaging resolution analysis that is consistent with theoretical predictions. Finally, concluding remarks are provided in Section IV.

## II. Problem Statement

### A. Wi-Fi based imaging setup

With reference to Fig. 1, the proposed Wi-Fi signal based through-the-wall imaging system consists of three main components: a conventional oscilloscope, Wi-Fi routers, and two receiving horn antennas. In our experiments, the scene of interest is located behind a 6cm-thick wooden wall. Several commercially available Wi-Fi routers (Mercury™ MW150R) are randomly deployed behind the wall. The Wi-Fi router utilized in our experiments is based on the IEEE 802.11n protocol, which works at 2.4GHz with 20MHz bandwidth. Two horn antennas are connected to the two ports of the oscilloscope (Agilent™ MSO9404A) and used to acquire the Wi-Fi signals scattered from the scene of interest. One receiving antenna is kept fixed and used for providing the reference signal. The other receiving antenna, which is mounted on a scanning platform, is controlled by a motor such that it can be moved along the *x*- and *y*- directions. The moving track follows an S-type motion. For each position of the scanning antenna along the track, the oscilloscope collects the data from the two antennas. In this way, a virtual two-dimensional antenna array can be formed akin to a synthetic aperture radar [2, 5].

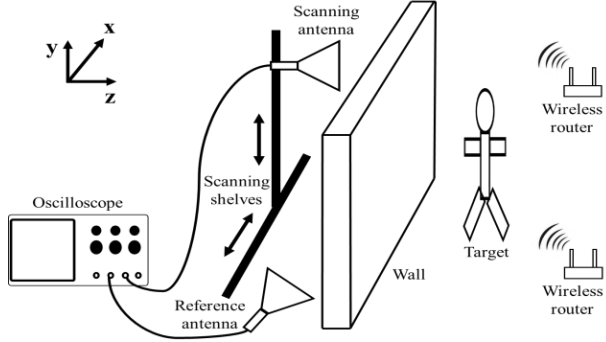

Fig.1. Schematics of the experimental setup.

*B. Data Processing*

We denote the time-domain Wi-Fi signal received by the reference antenna as $I_{ref}(t)$ and by the scanning antenna at a given location as $I_{sca}(t)$. The frequency-domain correlation function $R_{cov}$ of these two signals (hologram) is given by:

$$R_{cov} = \mathcal{F}^*\{I_{ref}(t)\} \times \mathcal{F}\{I_{sca}(t)\} \quad (1)$$

where $\mathcal{F}$ demotes a windowed Fourier transform. In our implementation, the width of time window of $\mathcal{F}$ is taken as 1μs, corresponding to the length of a whole Barker code. Figure 2 reports the normalized magnitude (top row) and phase (bottom row) of $R_{cov}$ at 2.4372GHz for three different investigated scene: no target (left column), metallic cross-shaped target as seen in Fig. 5 (mid column), and seated human subject as seen in Fig. 7a (right column). We use the classical filtered back-projection imaging algorithm [2, 5] to produce the images. The final (normalized) images as presented here are obtained as

$$B_n = \frac{B_o - B_b}{B_b} \quad (2)$$

where $B_o$ denotes the filtered back-projection image with the object present and $B_b$ denotes the back-projection data without the object.

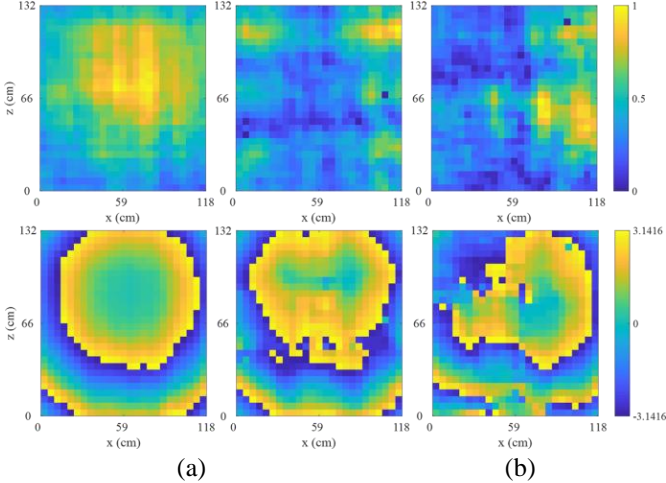

Fig. 2. The normalized magnitude (top row) and the phase (bottom row) of $R_{cov}$ at 2.42GHz for three different scenes: no target (left column), metallic cross-shaped target (mid column), and seated human subject (left column). For the latter two targets, see also Fig. 5 and Fig. 7a.

### III. RESULTS AND DISCUSSION

This section presents examples to demonstrate the performance of proposed Wi-Fi based imaging system. For the experiments considered in III.A and III. B, a Wi-Fi router is randomly placed behind the wall, as illustrated in Fig. 3(a).

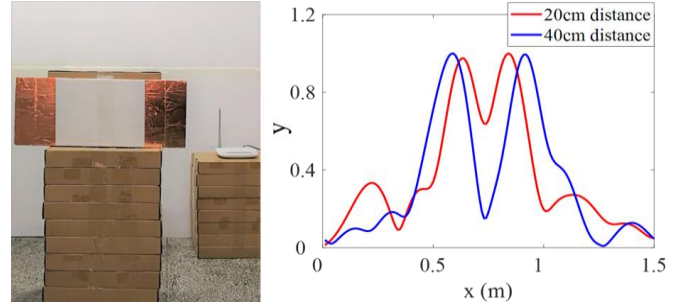

(a) (b)

Fig. 3. (a). Experimental setup with two separate metallic rectangular objects. (b). One-dimensional images of the objects separated by a distance of about 20cm and 40cm respectively. In this plot the y-axis represents the image normalized by the peak image intensity.

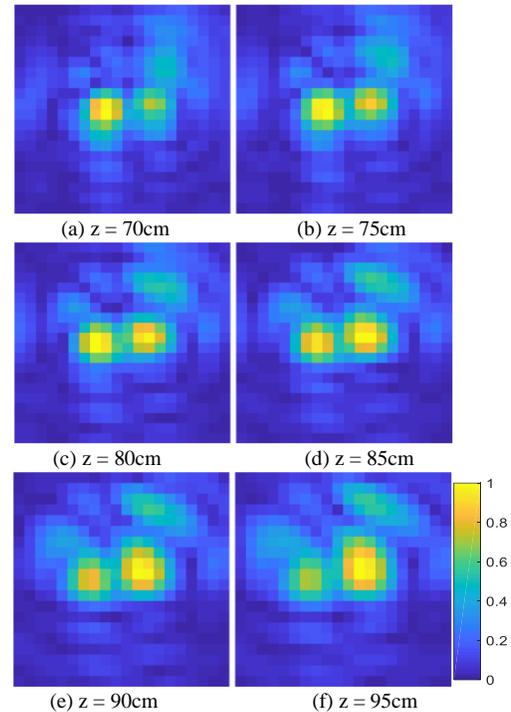

(a) z = 70cm   (b) z = 75cm
(c) z = 80cm   (d) z = 85cm
(e) z = 90cm   (f) z = 95cm

Fig. 4. Images of the two metallic objects as shown in Fig. 3(a) and separated by 10 cm along the x direction. The images are constructed along the *xy* plane for different depths (i.e. as *z* is varied, see also Fig. 1).

*A. Imaging of two rectangular metallic objects*

First, we consider two metallic objects placed about 20cm and 40cm apart of each other along the *x* direction respectively, as exemplified in Fig. 3(a). The scanning antenna moves from the left to the right along a 1m line, in intervals of 5cm, to provide 1-D imaging data. Fig. 3(b) shows the respective images, where the y-axis denotes the image normalized by the peak intensity. As indicated, the blue curve corresponds to the case of two objects with 40 cm separation and the red curve to the 20 cm separation. The separations between the two crests visible in these plots are consistent with the actual separation distances.

We next utilize the same two metallic objects to examine the depth resolution of our system. In this case, the separation of the two objects is fixed at about 10cm in the *x* direction. We perform filtered back-projection to obtain cross-plane (*xy* plane) images at different depths (i.e., as *z* is varied, see Fig. 1). These results are shown in Fig. 4. As expected, the best focusing occurs around the true location, which in this case is *z*=80cm. Fig. 4 shows that the two objects can be separated reasonably well for a cross range separation of 10cm.

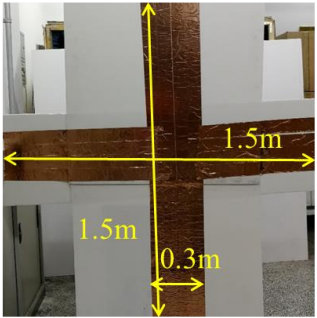

Fig. 5 the photo of metallic cross-shaped object considered in III. B

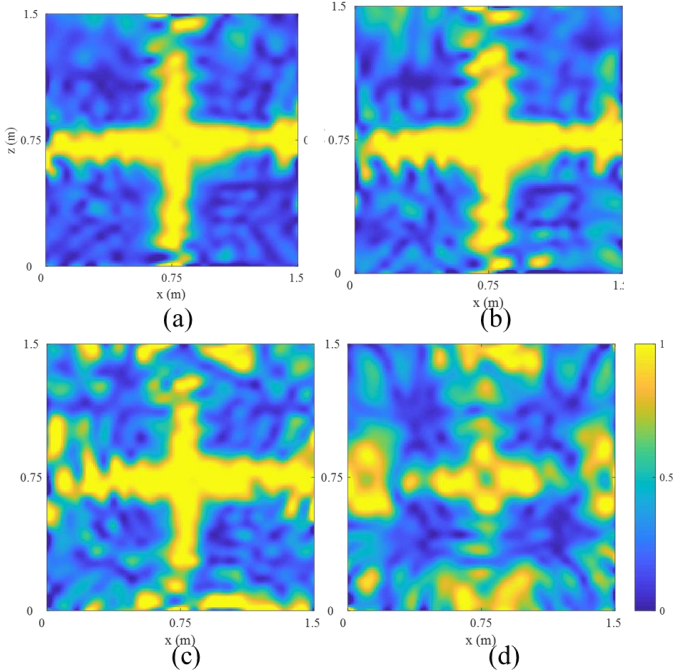

Fig. 6. The images of metallic cross-shaped object shown in Fig. 3(a) under different number of measurements. (a) the Nyquist sampling, (b) 1/2-sub Nyquist sampling, (c) 1/3-sub Nyquist sampling, and (d) 1/4-sub Nyquist sampling.

### B. Imaging of metallic cross-shaped object

In this section, we examine the imaging performance for a metallic cross-shaped target. The effects of different number of measurement acquisitions on the image quality is also investigated. The metallic cross is 1.5m long and 1.5m wide as shown in Fig. 5. The track of the scanning antenna spans approximately 117cm in the *x* direction and 130cm in the *y* direction. Fig. 6 shows the results for different number of measurements. In particular, Fig. 6(a) corresponds to Nyquist sampling and Figs. 6(b)-(d) correspond to 1/2-, 1/3-, and 1/4-sub-Nyquist sampling, respectively. With the exception of the lowest sampling, the images recover reasonably well the object shape, demonstrating the effectiveness of the 2-D Wi-Fi imaging system. Although this is not pursued here, it is expected that the image quality of sub-Nyquist sampling measurements can be improved by using sparsity-promoted reconstruction algorithms [2].

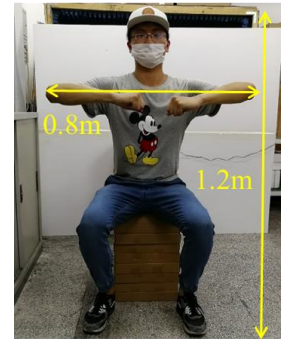

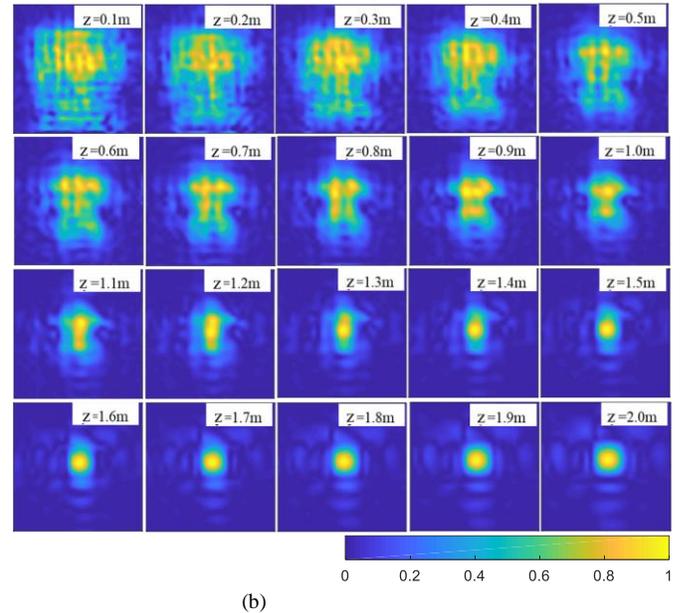

Fig. 7. (a) Seated human subject. (b) Corresponding images at the different slices along the *z* direction. A single Wi-Fi router is present in the environment.

### C. 3-D Imaging of human body

In this experiment, we consider the through-the-wall imaging of a human subject seated on top of a cardboard box.

First, we consider a case where a single Wi-Fi router is located at a distance of 1.3m behind (i.e. along *z* direction, see Fig. 1) the human subject as illustrated in Fig. 6(a). The mechanical scanning platform is at a distance 0.5m away from the subject along the *z* direction. Other setup parameters are the same as those used in Section III B. The results are shown in Fig. 7(b), where two-dimensional images at different depths along the *z* direction in increments of 10cm are provided. From these results, one can see that the best focused image is observed around the true range distance of 0.7m. Moreover, with the growth of the distance further away from the scanning platform, the image is gradually focused at the location of the

Wi-Fi router.

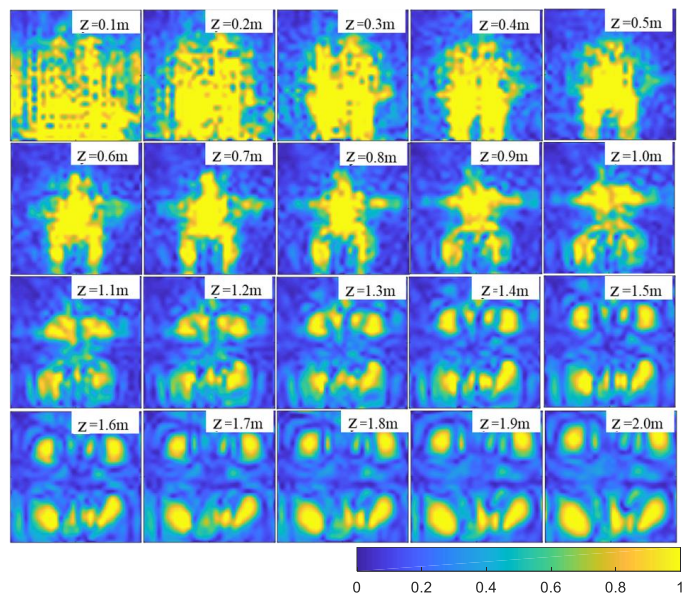

Fig. 8. The images of the human subject in Fig. 7(a) obtained at different slices along the *z* direction. In this case, four Wi-Fi routers are present in the environment.

Finally, we consider a case where four Wi-Fi routers are located 0.7m behind the seated human subject, at coordinates (*x*,*y*,*z*) = (0.3, 0.3, 2), (0.3, 1.2, 2), (1.2, 0.3, 2), and (1.2, 1.2, 2) in meters. Moreover, the wall and the human are located at 0.2m and 0.7m, respectively, along the *z* direction. Other parameters are the same as above. The associated images are shown in Fig. 8, where two-dimensional slices at different depths along *z* direction are compared. One can see that the best focused image of a human body can be observed again at the true range distance of 0.7m. Moreover, with the growth of the distance away from the scanning platform, the image is gradually focused towards the four Wi-Fi routers.

## IV. Conclusion

In this work, we experimentally verified that signals based on commodity 2.4GHz Wi-Fi devices can be exploited to realize three-dimensional through-the-wall imaging from the perspective of holographic imaging. In our implementations, a pair of antennas was used in a synthetic aperture mode to acquire the Wi-Fi signals scattered from the scene located behind a 6-cm thick wooden wall. Selected experimental results have been provided to investigate the performance of the proposed Wi-Fi based imaging scheme with respect to the number of routers. A simple filtered back-propagation algorithm was employed to form the images. More specialized reconstruction algorithms could be implemented to further improve the image quality and to extend this approach to thicker and/or denser walls.

Acknowledgement:
We would like to thanks Prof. Fernando L. Teixeira for insightful discussions.